\documentclass[
reprint,
amsmath,
amssymb,
aps,
prx,
superscriptaddress,
longbibliography,
nobalancelastpage,
]{revtex4-2}

\usepackage{graphicx}
\usepackage{dcolumn}
\usepackage{bm}
\usepackage[svgnames]{xcolor}

\usepackage{hyperref}
\usepackage{float}

\usepackage{physics2}
\usephysicsmodule[tightbraces=true]{ab}
\usephysicsmodule{braket,doubleprod}

\usepackage{comment}

\usepackage[separate-uncertainty=false,per-mode=symbol,print-unity-mantissa = false]{siunitx}
\DeclareSIUnit\gauss{G}
\DeclareSIUnit\centimeter{cm}
\DeclareSIUnit\barn{b}

\usepackage{braket}

\usepackage{editing}

\newchange{AA}{RoyalBlue}{Orange}
\newchange{ZZ}{CornflowerBlue}{Sienna}
\newchange{KM}{ForestGreen}{Sienna}
\newchange{MM}{ForestGreen}{Magenta}
\newnote{AA}{BlueViolet}
\newnote{ZZ}{CornflowerBlue}
\newnote{KM}{Olive}
\newnote{MM}{Goldenrod}
\changeAAShowNewOld
\noteAAShowFalse
\noteZZShowFalse
\noteMMShowFalse

\noteAllShowTrue

\usepackage{hyperref}

\makeatletter
\def\textemdash{\leavevmode\unskip\kern1pt---\kern1pt\ignorespaces}
\let\old@Section@Cmd=\section
\def\end@of@sec@tit{.\unskip---\kern1pt\ignorespaces}%
\def\mysection{%
  \def\reserved@stsec##1{\@startsection{section}{1}{\parindent}{\z@}{-0pt}{\normalfont\normalsize\itshape}*[##1]{##1\end@of@sec@tit}}%
  \@ifstar{%
    \reserved@stsec%
  }{%
    \reserved@stsec%
  }%
}
\let\section=\mysection
\makeatother

\hiddenblock{%
\def\refapp#1{Appendix~\ref{#1}}%
}{%
\def\refapp#1{Ref.~\cite{SuppMat}}%
}

\newcommand{\figref}[2]{\hyperref[#1]{\ref{#1}(#2)}}

\def\Title{A cryogenic neutral-atom platform with full optical access and 2-hour trap lifetime}

\hypersetup{
  pdftitle = {\Title},
  pdfauthor = {MQV atoms team, MQV cryo team, and  Immanuel Bloch},
  colorlinks = true,
  linkcolor = blue,
  citecolor = blue,
  filecolor = blue,
  urlcolor = blue,
}

\def\LMU{Fakultät für Physik, Ludwig-Maximilians-Universität München, 80799 München, Germany}
\def\MCQST{Munich Center for Quantum Science and Technology, 80799 München, Germany}
\def\MPQ{Max-Planck-Institut für Quantenoptik, 85748 Garching, Germany}
\def\PLANQC{PlanQC GmbH, 85748 Garching, Germany}

\begin{document}

\title{\Title}

\author{Akhil Kumar}
\affiliation{\MPQ}
\affiliation{\LMU}
\affiliation{\MCQST}

\author{Lorenzo Festa}
\affiliation{\MPQ}
\affiliation{\MCQST}

\author{Avishay Grinberg}
\affiliation{\MPQ}
\affiliation{\LMU}
\affiliation{\MCQST}

\author{Eran Reches}
\affiliation{\MPQ}
\affiliation{\LMU}
\affiliation{\MCQST}

\author{Dimitrios Tsevas}
\affiliation{\MPQ}
\affiliation{\LMU}
\affiliation{\MCQST}

\author{Kevin P. Mours}
\affiliation{\MPQ}
\affiliation{\LMU}
\affiliation{\MCQST}

\author{Zhao Zhang}
\affiliation{\MPQ}
\affiliation{\LMU}
\affiliation{\MCQST}

\author{Robin Eberhard}
\affiliation{\MPQ}
\affiliation{\LMU}
\affiliation{\MCQST}

\author{Sebastian Blatt}
\affiliation{\MPQ}
\affiliation{\LMU}
\affiliation{\MCQST}
\affiliation{\PLANQC}

\author{Andrea Alberti}
\affiliation{\MPQ}
\affiliation{\LMU}
\affiliation{\MCQST}

\author{Johannes Zeiher}
\affiliation{\MPQ}
\affiliation{\LMU}
\affiliation{\MCQST}

\author{Immanuel Bloch}
\affiliation{\MPQ}
\affiliation{\LMU}
\affiliation{\MCQST}

\author{Max Melchner}\email{max.melchner@mpq.mpg.de}
\affiliation{\MPQ}
\affiliation{\LMU}
\affiliation{\MCQST}

\date{\today}

\begin{abstract}

Neutral-atom quantum processors are rapidly scaling toward system sizes of more than ten thousand qubits, allowing for the realization of a new class of quantum computing algorithms and quantum simulation experiments. However, current neutral-atom platforms generally have to find a compromise between the optical accessibility and the storage time of atoms in optical potentials, limiting the available qubit numbers. Here we report on the operation of a novel, cryogenically enhanced, neutral-atom apparatus that overcomes these apparently conflicting requirements. We demonstrate vacuum-limited trapping lifetimes of up to two hours of single $^{88}\mathrm{Sr}$ atoms in an optical tweezer array while preserving full optical access and without the need for complex cryogenic enclosures. Our measurements show that exceptionally long single-atom lifetimes can be achieved with a relatively simple cryostat design. Our architecture can be straightforwardly ported to other atomic species and shows a viable path for scaling up to sorted arrays of tens of thousands of atoms.

\end{abstract}

\maketitle

\section{Introduction}

\begin{figure*}[htb]
    \centering
    \includegraphics{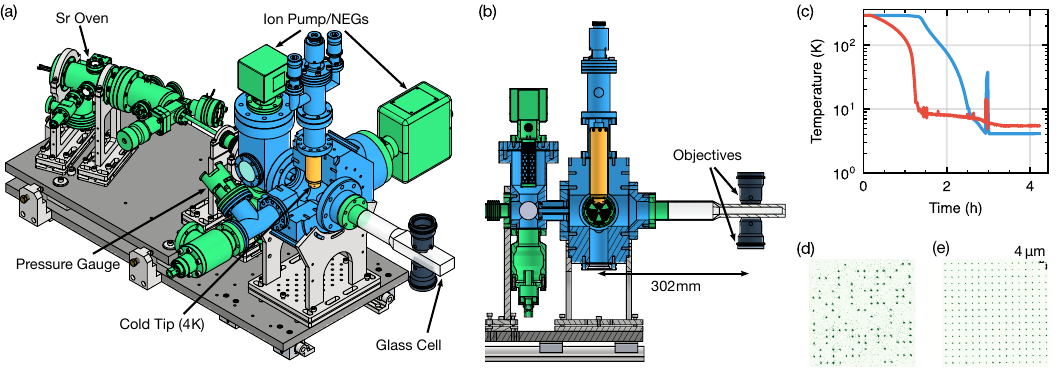}
    \caption{Render of the vacuum chamber. (a) The vacuum chamber consists of a strontium oven, a chamber made predominantly of grade 2 titanium, a glass cell, and a cryostat that cools a cold tip in vacuum to \SI{4}{\kelvin}. The components are colored according to the material of the vacuum-facing surfaces. Titanium, stainless steel, and copper surfaces are colored in blue, green, and orange, respectively. Atoms are trapped in optical tweezers in a fused silica glass cell. The objectives used to focus the tweezers into the vacuum chamber and collect fluorescence for tweezer-resolved imaging are shown in black. (b) Cross section of the right part of the chamber. The cold tip is located roughly \SI{30}{\centimeter} away from the tweezer array. A typical cool-down protocol to cryogenic temperatures is shown in (c). The cold tip is cooled to \SI{4}{\kelvin} from room temperature approximately three hours after the helium compressor circuits are started. The temperatures of the cryocooler cold tip and cryostat cold tip are shown in red and blue, respectively. The temperature spike at the third hour occurs during removal of a cold trap from the helium circuit, when the circuit is partially closed. The temperature quickly drops to \SI{4}{\kelvin} after the cold trap is removed and the helium circuit is reopened. (d) A single-shot tweezer image. (e) Averaged image of the $16\times 16$ tweezer array. The spacing between neighboring tweezers is \SI{4}{\micro\meter}.} 
    \label{fig:1}
\end{figure*}

Trapping and manipulating neutral atoms in optical potentials is a key technology for quantum simulation and quantum computation \cite{Bloch2008, Bloch2012, Saffman2016}. In quantum simulators, ultracold atoms probe the ground-state properties and nonequilibrium dynamics of engineered many-body systems \cite{Dalfovo1999, Jaksch2005, Lewenstein2007, Gross2017}.
In neutral-atom quantum computers, individually trapped atoms form quantum registers, and quantum gates are driven with lasers or microwave radiation \cite{Jaksch2000, Saffman2010, Endres2016}.

As the field of neutral-atom quantum processing advances, the system sizes are increasing. Recently, sorted and unsorted arrays of several thousands of atoms have been realized. Several technologies and techniques have enabled the creation of large \cite{Bakr2009, Sherson2010, Wang2020, Impertro2023, Tao2024, Wang2026}, ordered atom arrays \cite{Pichard2024}, including large-field-of-view objectives \cite{Manetsch2025, Chiu2025}, continuous reloading \cite{Norcia2024, Gyger2024, Li2025, Chiu2025}, and SLM sorting \cite{Knottnerus2025, Lin2025, Huang2026}.

To increase the system size, it is necessary to keep more atoms trapped for the duration of sorting, and during the experiment. This becomes more challenging as the system size increases \cite{Barredo2016,  Sheng2021}, since the time it takes to sort generally scales linearly with the number of atoms and the probability of preserving a defect-free array decreases exponentially with array size. 

One limitation on the achievable number of atoms in an ordered array is their lifetime in optical traps. In addition to constraints related to the trap, such as heating caused by intensity or pointing noise, the lifetime is determined by the vacuum pressure in the vacuum chamber. The vacuum pressure in ultra-high-vacuum (UHV) chambers is established through an equilibrium between gas sources, such as diffusion into the vacuum chamber from the outside and outgassing from chamber walls, and pumps, such as ion and getter pumps. Hydrogen outgassing from chamber walls is generally considered to be the dominant source of residual gas load in UHV chambers \cite{Redhead2003}. 

Several methods have been proposed to lower the pressure in vacuum chambers used for neutral-atom experiments. For example, cryogenic surfaces inside the vacuum chamber can act as a cold trap for residual background gases \cite{Baglin2020}. The design of these cryogenic apparatuses has generally been quite complex, balancing the desire to enclose as much of the atomic array in a \SI{4}{\kelvin} environment while simultaneously allowing just enough optical access to manipulate the atoms. This tradeoff has meant that while state-of-the-art cryogenic apparatuses demonstrate significant improvements in vacuum-limited lifetime, and some success in suppressing room-temperature blackbody radiation \cite{Jin2026, Cantat2020}, they are generally limited in their capabilities, due in part to the lack of optical access, and the need to use in-vacuum windows \cite{Schymik2021, Schymik2022, Pichard2024, Hassan2025, Zhang2025, Lim2026}.  

Here, we report on a cryogenically-enhanced neutral-atom tweezer machine that overcomes the tradeoff between lifetime and optical access. Our machine sets a record for the longest vacuum-limited lifetime observed for neutral atoms in optical potentials ($>$\SI{7000}{\second} or $\sim$\SI{2}{\hour}), while simultaneously providing maximum optical access. This advance is the result of (1) a specific choice of materials for the vacuum apparatus (2) an apparatus design that allows the UHV chamber to be baked, and (3) a cold tip at \SI{4}{\kelvin} inside the vacuum chamber, that pumps residual hydrogen outgassing from vacuum-facing chamber walls. 

\section{Apparatus Design}

\begin{figure*}[htb] 
    \centering
    \includegraphics{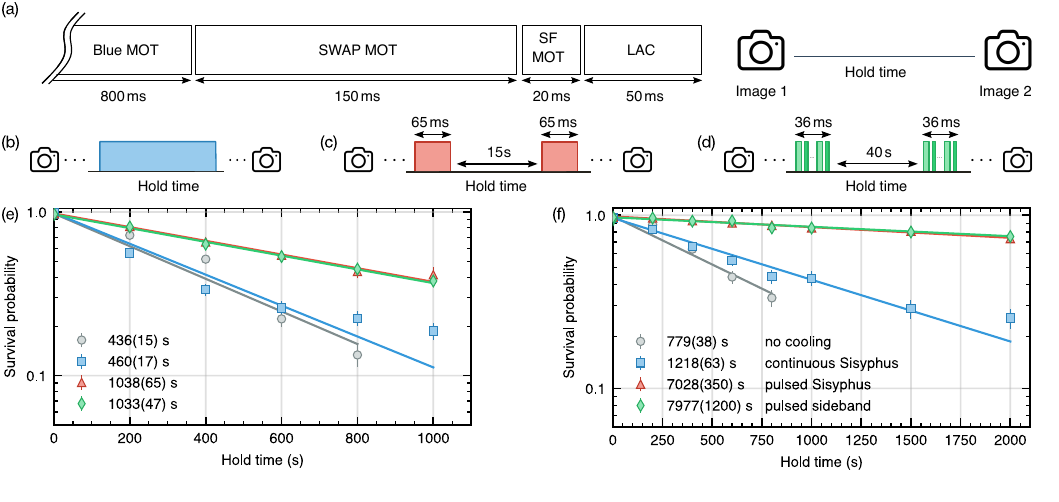}
    \caption{Lifetime measurements in the cryogenic apparatus. (a) A schematic of our experimental sequence. Following state preparation and initial imaging, single atoms are held in optical tweezers for a variable duration (the “hold time”). During this hold time, we apply either continuous Sisyphus cooling (b), pulsed Sisyphus cooling (c), or pulsed sideband cooling (d) and measure the corresponding atomic lifetime. Continuous Sisyphus cooling is maintained throughout the entire hold period. For pulsed Sisyphus cooling, we apply individual cooling pulses every \SI{15}{s} for a duration of \SI{65}{ms}. Our pulsed sideband cooling protocol consists of alternating pulses of radial and axial sideband cooling with a total duration of \SI{36}{ms} every \SI{40}{s}. Panel (e) and (f) show the measured survival probability of single atoms confined in optical tweezers as a function of the hold time when the cryostat is disabled and enabled, respectively. Different cooling protocols during the hold time, i.e., no cooling, continuous Sisyphus cooling, pulsed Sisyphus cooling and pulsed sideband cooling are shown in grey, blue, red and green, respectively. The atomic lifetime in each case is extracted by fitting the survival probability to an exponential decay model. The difference in atomic lifetime between the situation in which the cryostat is disabled and enabled highlights the impact of improved vacuum conditions on the trapped atom lifetime, as well as the interplay between background gas collisions and trap-induced heating under different cooling configurations. From the differential AC Stark shift between the $5\text{s}^2\, {}^1\text{S}_0$ and $5\text{s}5\text{p}\, {}^3\text{P}_1$ levels, we estimate the average trap depth of the tweezers to be around \SI{220}{\micro K}, with a standard deviation in trap depth of $\sim$2$\%$ (Appendix D).}
    \label{fig:2}
\end{figure*}

A render of our vacuum chamber is shown in Fig.~\ref{fig:1}. We combine a commercial strontium oven (AOSense) with a bakeable UHV chamber, a glass cell, and a commercially available closed-cycle cryostat (ColdEdge Stinger, procured through Cryoandmore) that can be cooled to \SI{4}{\kelvin}. We ensured that as many vacuum-facing surfaces as possible were manufactured from grade 2 titanium. The outgassing characteristics of grade 2 titanium, especially with regard to hydrogen, are superior to stainless steel, a common material used to construct vacuum chambers \cite{Ishizawa2006, Fedchak2021}. Titanium sublimation pumps are also often used as evaporable getter pumps in UHV environments \cite{Benvenuti2007}. 

In Fig.~\ref{fig:1}, components made of grade 2 titanium are colored in blue, while those made of stainless steel, grade 304 or 316, are shown in green. The vacuum-facing surfaces of the cryostat itself are either OFHC (Oxygen-Free High Conductivity) copper or grade 2 titanium.

To further reduce outgassing, the vacuum chamber went through two bakeouts, one `high-temperature' bakeout of the core titanium chamber without the glass cell at an average temperature of $320^\circ\mathrm{C}$ for 3 weeks, and one `low-temperature' bakeout of the full chamber at an average temperature of ~$150^\circ\mathrm{C}$ for 2 weeks.

For our experiments, we load $^{88}\mathrm{Sr}$ atoms into a $16\times 16$ square optical tweezer array formed by an \SI{8}{W} laser at 813 nm (Precilasers, FL-SF-813-8-CW) and a spatial light modulator (Hamamatsu, X15213-02R LCOS-SLM). The tweezer spacing is \SI{4}{\micro m}. The diffracted light from the SLM is magnified and focused onto the atomic plane with an $\text{NA} = 0.65$ objective (SpecialOptics). Atomic fluorescence is detected via the same objective.

Unlike enclosed cryostats, where imaging optics are generally placed in vacuum and designed for cryogenic operation, and thus have to be compatible with thermal contraction and UHV conditions, our design puts all objectives outside the glass cell, enabling the use of standard commercial optics operating at room temperature. One important caveat of our design is that it does not shield the atoms from exposure to room-temperature blackbody radiation.

Remarkably, activating the cryostat leads to a pronounced reduction in vacuum pressure at the center of the glass cell, despite a separation of approximately \SI{30}{\centimeter} between the tweezer array and the cold tip. This indicates that the vacuum pressure can still be lowered deep in the molecular-flow regime of our UHV chamber and that line-of-sight exposure to the dominant outgassing load, the Sr oven, is not necessarily a hard limitation to the vacuum pressure and vacuum-limited lifetime. The observed decrease in pressure upon activating the cryostat further indicates that the residual gas is dominated by hydrogen rather than helium diffusing through the glass cell \cite{Norton1953}, since the cold tip does not effectively pump helium.

\section{Enhanced lifetime}

A schematic of the experimental sequence used to measure the lifetime of atoms in our optical tweezers is shown in Fig.~\ref{fig:2}(a)-(d). Atoms are first loaded into the tweezer array using a two-stage cooling process consisting of a \SI{800}{ms} blue magneto-optical trap (MOT), followed by a \SI{150}{ms} SWAP MOT stage \cite{Snigirev2019}, and \SI{20}{ms} single-frequency (SF) red MOT to further cool and compress the atomic cloud. After loading, a parity-projection pulse is applied, which deterministically removes atom pairs via light-assisted collisions (LAC) and leaves each tweezer occupied by either zero or one atom. We take an initial fluorescence image to determine the spatial distribution and occupancy of atoms across the tweezer array. Following this, the atoms are held in the optical tweezers for a variable duration (the “hold time”). After this hold time, a second image is acquired to measure the number of atoms that remain trapped, allowing us to extract the survival probability as a function of hold time.

To investigate the factors limiting the atomic lifetime, we perform this sequence under several conditions during the hold time: no cooling, continuous and pulsed Sisyphus cooling \cite{Cooper2018}, and pulsed sideband cooling \cite{Norcia2018}. By comparing these cases, we can distinguish the effects of heating from the trapping light and background gas collisions, and assess how different cooling strategies influence the lifetime of atoms in the optical tweezers.

We measure a vacuum-limited atomic lifetime of over \SI{7000}{s} (approximately \SI{2}{hours}) when applying pulsed Sisyphus cooling and pulsed sideband cooling on the  $5\text{s}^2\, {}^1\text{S}_0$--$5\text{s}5\text{p}\, {}^3\text{P}_1$ transition, as shown in Fig.~\ref{fig:2}. In addition, we measure lifetimes of several hundred seconds even in the absence of cooling, which we attribute to the vacuum pressure already being in the \SI{1e-11}{\milli\bar} regime without running the cryostat. 

We see a gradual degradation of the vacuum quality due to the adsorption of hydrogen on the \SI{4}{\kelvin} surfaces, similar to what has been observed elsewhere \cite{Pichard2024, Zhang2025, Lim2026}. From long-term measurements, we extract a characteristic $1/e$ decay time of approximately \SI{10}{days} for the lifetime, indicating a gradual increase in background pressure over time (Appendix A). 

To further probe the role of hydrogen, we deliberately heat the cryogenic surface (cold tip) to \SI{30}{\kelvin}, resulting in a pronounced pressure spike consistent with the release of adsorbed hydrogen. After the cold tip is subsequently cooled back to \SI{4}{\kelvin}, the vacuum conditions are restored, and we again observe atomic lifetimes beyond \SI{7000}{s}. This behavior strongly supports the conclusion that the cold tip acts as an efficient cryopump for hydrogen, the dominant residual gas that limits the lifetime of our system at room temperature.  

\begin{figure}[htb] 
    \centering
    \includegraphics[scale=1.0]{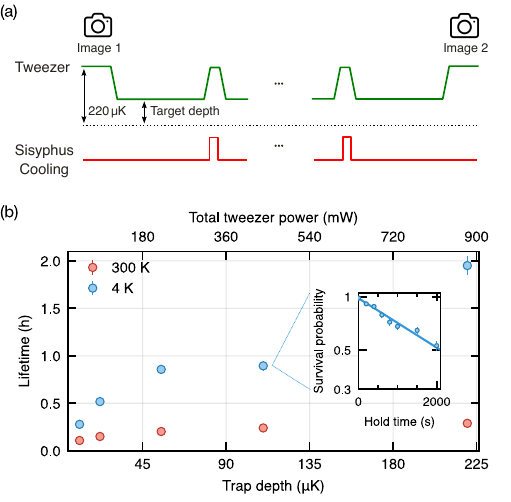}
    \caption{Measuring lifetime as a function of tweezer depth. (a) Experimental sequence for measuring the trapped-atom lifetime at different tweezer powers. After the initial fluorescence image, the tweezer depth is linearly ramped down to a target trap depth in \SI{5}{ms}. The trap depth is ramped up to 100\,\% for each Sisyphus cooling pulse; the trap depth is thus higher than the target value for a total duration equal to 0.5\,\% of the hold time.  Measurements are performed in both cryogenic and non-cryogenic environments. (b) Lifetime vs tweezer trap depth with cryopumping enabled (blue) and disabled (red). Each lifetime value is extracted from a fit to an exponential decay of the atom survival vs hold time in the tweezer (inset). The total tweezer power quoted here is equal to the power per tweezer inside the glass cell multiplied by the total number of tweezers (Appendix D).}
    \label{fig:3}
\end{figure}

\section{Lifetime vs trap power and oven temperature}

In cryogenic neutral-atom platforms, the total optical power delivered to the vacuum chamber constitutes an important systems-level constraint, particularly in the context of scaling to large tweezer arrays. Unlike room-temperature setups, conventional \SI{4}{\kelvin} cryostats operate with a finite cryogenic cooling power margin, such that absorption and scattering of several watts of near-infrared trapping light can lead to measurable heating of the cryogenic shields and surrounding structures. 

In recent work demonstrating thousand-atom defect-free arrays in a \SI{4}{\kelvin} environment \cite{Lim2026}, a reduction of the vacuum-limited atom lifetime with increased tweezer power was reported, attributed to a degradation of cryopumping efficiency and enhanced desorption of residual hydrogen from cryogenic surfaces. These observations highlight a potential scaling bottleneck in cryogenic tweezer architectures, where increasing array sizes typically require higher total optical power, thereby increasing the thermal load on the cryogenic system and potentially compromising vacuum conditions required for long trapping lifetimes and high-fidelity operation.

We directly investigate the dependence of the trapped-atom lifetime on the total optical power delivered to the cryogenic vacuum chamber by lowering the tweezer power to a target value between the imaging and cooling pulses (Fig.~\ref{fig:3}). To ensure that our measurements are not affected by the gradual decay of the pumping efficiency of the cold tip, we regenerate the cold tip before every new lifetime measurement (see Appendix A). 

We find that the lifetime of atoms increases as we increase the depth of our tweezers in both the room-temperature and the cryogenic configuration. In addition, we do not see a change in the temperature of the cold tip as we increase the tweezer power. The large distance between the \SI{4}{\kelvin} surface and the tweezer array ensures that the cold tip is not affected by direct or stray exposure to the tweezer light. The increase of the lifetime with tweezer power is likely due to atoms having to overcome a higher potential barrier to escape the trap and lower relative intensity noise of the tweezer light at higher powers.

The absence of a measurable dependence of the temperature of the cryogenic surface on total optical power, together with exceptionally high vacuum-limited lifetimes, is a key to scaling up the atom number. Our cryogenic platform can accommodate the increased optical power requirements associated with large-scale tweezer arrays without compromising vacuum conditions or atom retention.

In a separate set of measurements, we investigate the dependence of the trapped-atom lifetime on oven temperature. As shown in Fig.~\ref{fig:4}, the lifetime decreases monotonically with increasing oven temperature, consistent with an elevated background gas pressure in the trapping region. By contrast, the lifetime exhibits no significant dependence on the MOT loading time, indicating that Sr flux is not the dominant loss mechanism. This points instead to oven outgassing, likely hydrogen, as the primary source of atom loss and identifies the oven temperature as a key parameter governing the lifetime in our apparatus. A straightforward chamber modification that eliminates direct line-of-sight exposure between the oven and the glass cell could further improve performance at higher oven temperatures.

{
\setlength{\belowcaptionskip}{-10pt}
\begin{figure}[h]
    \centering
    \includegraphics[scale=1.0]{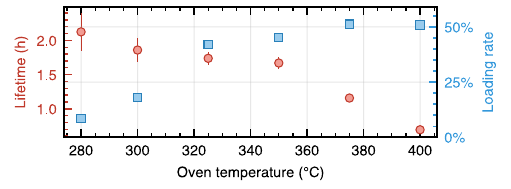}
    \caption{Measuring lifetime as a function of oven temperature. We observe a pronounced dependence of the trapped-atom lifetime on the oven temperature, with lower oven temperatures consistently leading to longer trapping lifetimes. This behavior is attributed to the reduced outgassing of hydrogen from the oven and, correspondingly, a lower background collision rate at lower oven temperatures. At the same time, decreasing the oven temperature also reduces the atom loading rate into the tweezers, reflecting the trade-off between loading efficiency and vacuum-limited lifetime. Measurements use a blue MOT loading time of \SI{800}{\milli\second} and a tweezer depth of \SI{220}{\micro\kelvin}. Lower loading rates can be partially compensated by longer loading times.}
    \label{fig:4}
\end{figure}
}

\section{Conclusion and Outlook}

Here we demonstrate that long single-atom lifetimes can be achieved with a relatively simple cryostat. The combination of a bakeable grade 2 titanium chamber with a simple cold finger at \SI{4}{\kelvin} allowed us to measure a vacuum-limited lifetime on the order of two hours. Furthermore, this chamber design allows for long lifetimes even with the cold tip at room temperature, and even without having to cool the atoms. The latter should allow us to trap the long-lived metastable states of $^{88}\mathrm{Sr}$, i.e., the $5\text{s}5\text{p}\, {}^3\text{P}_0$ and $5\text{s}5\text{p}\, {}^3\text{P}_2$ states, which are not straightforwardly coolable, for longer than was previously possible \cite{Dolde2025, Kim2025}. 

A straightforward upgrade to the cryostat is to substantially increase the surface area cooled to \SI{4}{\kelvin}, e.g. by attaching a shield to the cold tip and/or introducing activated charcoal. This should both increase the vacuum-limited lifetime and increase the time before hydrogen adsorption limits the cryo pumping effect. We anticipate that a properly designed shield should allow for a roughly 100-fold increase in surface area at \SI{4}{\kelvin}.

Our cryogenic design substantially improves the vacuum-limited lifetime, but does not provide significant shielding from blackbody radiation because the tweezer array remains exposed to room-temperature surfaces. This contrasts with fully enclosed cryogenic platforms, where blackbody radiation can be strongly suppressed \cite{Hassan2025, Jin2026, Cantat2020}. However, our design remains fully compatible with room-temperature microwave shielding \cite{Meinert2020, Holzl2024}, offering a potential route to enhance Rydberg state lifetimes alongside long storage times.

Finally, with the ability to trap atoms longer, the obtainable system size also naturally increases. The longer atoms are trapped, the more time can be spent sorting them into regular arrays. We expect that our apparatus will allow for the assembly of defect-free atom arrays of several tens of thousands of atoms with high probability. Our full optical access makes it possible to install large-field-of-view objectives outside the vacuum chamber and create large-volume optical lattices with high-power lasers~\cite{Tao2024, Gyger2024}, providing a realistic path towards scaling atom arrays to beyond 100,000 atoms.

\section{Acknowledgements}
We thank Íñigo Lasheras López-Cerón for his help building the vacuum chamber and for contributing to the assembly of various optical setups. We are also grateful to Milan Antic, Felix Friedrich, and Anton Mayer for their assistance in designing many parts of the apparatus. Special thanks goes to Ting You Tan, Zhenpu Zhang, and Adam Kaufman for insightful discussions.

We acknowledge funding by the Max Planck Society (MPG), the Deutsche Forschungsgemeinschaft (DFG, German Research Foundation) under Germany's Excellence Strategy--EXC-2111--390814868, and through JST-DFG2024: Japanese-German Joint Call for Proposals on “Quantum Technologies” (Japan-JST-DFG- ASPIRE 2024) under DFG Grant No. 554561799, from the Munich Quantum Valley initiative as part of the High-Tech Agenda Plus of the Bavarian State Government, and from the BMFTR through the programs MUNIQC-Atoms (Grant No. 13N16070)  and MAQCS (Grant No. 13N16895).
J.Z. acknowledges support from the BMFTR through the program “Quantum technologies---from basic research to market” (SNAQC, Grant No. 13N17337).
        
Competing interests: S.B. and J.Z. are co-founders and shareholders of PlanQC GmbH.

\appendix
\section{Appendix A: Regeneration process}
\label{sec:regeneration}

\begin{figure}[h!]  
    \centering
    \includegraphics[scale=1.0]{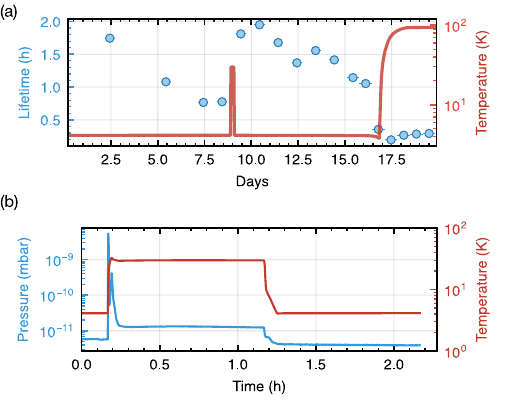}
    \caption{Regenerating the cold tip. (a) The lifetime is continuously monitored over 16 days by running the lifetime sequence described in the main text. Around day 9, the cold tip was regenerated by heating it up for a few hours. Around day 17, the temperature gradually increased to \SI{100}{\kelvin}. 
    As the temperature increases, hydrogen desorbs from the cryogenic surface at a higher rate and the lifetime decreases significantly. The lifetime recovers after a few hours, indicating the hydrogen released during the heat-up of the cold tip has been removed by other parts of the vacuum system (e.g., getters/ion pumps). After regenerating the cold tip, we measure an exponential decay of the lifetime with a time constant of \SI{10}{days}. A zoomed-in version of the regeneration process is shown in panel (b).}
    \label{fig:5}
\end{figure}

We find that the atomic lifetime gradually decreases over time from an initial value around \SI{2}{h}. Assuming that the lifetime reaches a steady-state value equal to the non-cryogenic case, we find a $1/e$ decay time of \SI{10}{days} (Fig.~\ref{fig:5}). To recover the initial lifetime, we heat the cold tip to \SI{30}{K} via heaters internal to the cryostat. The pressure accordingly spikes initially, indicating the outgassing of adsorbed hydrogen \cite{Redhead2003}, and then drops to a steady-state value of $\sim$\SI{1e-11}{\milli\bar} over time, equal to the pressure we measure at room temperature. After approximately \SI{1}{\hour}, we turn off the heaters and the cold tip quickly cools back down to \SI{4}{\kelvin}. As a result, the pressure drops to $\sim$\SI{2.5e-12}{\milli\bar}, close to the detection limit of the pressure gauge (Leybold IONIVAC IE 514).

\section{Appendix B: Stinger system}

The cryogenic part of the system consists of two interconnected, closed-cycle helium circuits. The first circuit supplies pressurized helium to drive a two-stage cryocooler (Sumitomo RDK-415D2), whose cold tip is cooled down to roughly \SI{5}{\kelvin} in normal operation. In the second helium circuit, driven by a second helium compressor (KNF 630N.15), the gaseous helium is cooled by the cryocooler and is directed through the cold tip of the cryostat on the high-pressure side of the vacuum chamber. This cools the vacuum-facing side of the cold tip to \SI{4}{\kelvin}. As the spent helium returns back to the helium compressor, it cools a radiation shield in the vacuum chamber to roughly \SI{20}{\kelvin}. The total surface area that is cooled to \SI{4}{\kelvin} is about \SI{40}{\centimeter\squared}.

We found that the cold tip of the Stinger system initially remained at \SI{4}{\kelvin} for roughly \SI{16}{\hour} before heating up to room temperature. Only replacing the helium in the secondary helium circuit allowed the cold tip to cool down to \SI{4}{\kelvin}. We decided to replace the compressor driving the secondary helium circuit with an oil-free alternative (KNF 630N.15). After this change, the cold tip stayed at \SI{4}{\kelvin} for more than 2 weeks.
It takes roughly half a day for the stinger to heat up to room temperature, the helium to be replaced, and the stinger to cool down to \SI{4}{\kelvin}. We anticipate that the time the cold tip remains at \SI{4}{\kelvin} can be improved by adding additional filters and sealing small leaks in the secondary helium circuit. 

To clean the helium in the secondary helium circuit, we let the helium flow through a cold trap during the 3-hour cooldown period. We remove the cold trap right after the cold tip reaches \SI{4}{\kelvin}. For the measurements presented in this work, we had one helium gas purifier (Supelco 27601-U) and one molecular sieve adsorber (Sumitomo 257428D) installed in the secondary helium circuit. 

\section{Appendix C: Data Analysis}

We extract lifetimes by fitting the survival probability (fraction of atoms remaining between the first and second images) versus hold time to a single exponential model $A \exp(-t/\tau)$, with $A$ and $\tau$ treated as free parameters in a nonlinear least-squares fit. The reported lifetimes were extracted from the fitted $\tau$, with statistical uncertainties obtained from the covariance matrix. The model assumes a single dominant decay channel over the fitted range, consistent with structureless residuals within experimental uncertainty.

\section{Appendix D: Determination of optimal cooling conditions} 
\begin{figure}[h!] 
    \centering
    \includegraphics[scale=1.0]{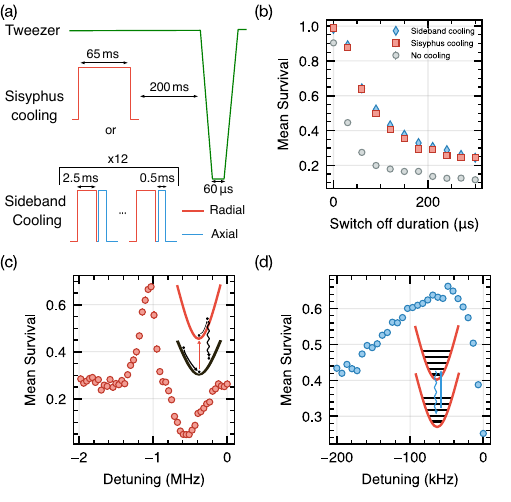}
    \caption{(a) Typical experimental sequence used to optimize the Sisyphus and sideband-cooling stages. A release-and-recapture measurement is performed after \SI{200}{ms} of a single cooling block to enhance the sensitivity of the optimization. (b) Atom survival probability as a function of tweezer switch-off duration, comparing the cases with and without cooling. (c), (d) Frequency dependence of the survival after \SI{60}{\micro\second} release-and-recapture for Sisyphus cooling and sideband cooling, respectively. The frequency is given as a detuning from the free-space $5\text{s}^2\, {}^1\text{S}_0$-$5\text{s}5\text{p}\, {}^3\text{P}_1$ transition. For the sideband cooling trace, the axial sideband cooling beam is red-detuned from the carrier transition by roughly four times the axial trap frequency, and the frequency of the radial sideband cooling beam is scanned.}
    \label{fig:6}
\end{figure}

We investigated several cooling schemes during the lifetime measurements and optimized their parameters to maximize performance under experimental conditions. To quantitatively assess and compare the effectiveness of the different cooling methods, we employed a release-and-recapture technique \cite{Tuchendler2008}. In this approach, atoms are briefly released from the trapping potential and subsequently recaptured. The recapture probability serves as a proxy for the temperature of the atoms in the tweezers and allows us to optimize the cooling parameters by maximizing the recapture probability as shown in Fig.~\ref{fig:6}. In particular, systematic scans were performed for both pulsed Sisyphus cooling and sideband cooling (pulse duration, detuning, and optical power). The resulting optimized settings were then used in the lifetime measurement sequence to ensure that the observed lifetimes were obtained under conditions of maximum cooling efficiency and reproducibility. 

We combine the value of the shift of the $5\text{s}^2\, {}^1\text{S}_0$-$5\text{s}5\text{p}\, {}^3\text{P}_1$ transition with our measured radial trap frequency of \SI{57}{kHz} to determine the tweezer waist and the power per tweezer. 

\clearpage
\bibliography{cryoreferences}
\end{document}